\documentclass[prb,twocolumn,showpacs,preprintnumbers,amsmath,amssymb]{revtex4}
\usepackage{graphicx}
\usepackage{dcolumn}
\usepackage{bm}
\usepackage[T1]{fontenc}
\usepackage[ansinew]{inputenc}
\usepackage{amsmath}
\usepackage{amssymb}
\usepackage{graphicx}
\usepackage{color}
\usepackage[colorlinks]{hyperref}

\begin{document}

\title{Comment on "A simple experiment to test Bell's
inequality", J.-M. Vigoureux.  }

\author{A.~Drezet}
 \affiliation{Institut f\"ur
Experimentalphysik, Karl Franzens Universit\"at Graz,
Universit\"atsplatz 5 A-8010 Graz, Austria}

\date{\today}
\begin{abstract}
In the above paper, it is claimed that with a particular use of the Bell inequality
a simple single photon experiment could be performed to show the
impossibility of any deterministic hidden variable theory in quantum
optics. A careful analysis of
the concept of probability for hidden variables and a detailed discussion of the hidden variable model of de Broglie-Bohm show that the reasoning and main conclusion of this paper are
not correct.

\end{abstract}
\maketitle

\section{Introduction}

It has been claimed in a recent article [1] that deterministic
hidden variable theories are not admissible in the context of
quantum mechanics. Following this paper ``it is not possible to
consider that the photon is programmed once for all from its
emission and that it possesses a flight plan indicating its
behaviour in any occasion.'' In order to prove this result, the
author has proposed a simple experiment using only single photon
events in order to show a contradiction with the Bell inequality
[2-4]. Would this result be correct, a return back to the situation
existing prior to the refutation by J.~S.~Bell [5] of the well known
theorem of Von Neumann [6] on the absence of dispersion-free states
would have been achieved (For a report that Einstein earlier raised
the same critics than Bell see [7] p.~89). However this theorem was
already refuted by de Broglie [8,9] and Bohm [10] on the simple fact
that the existence of their proper deterministic hidden variable
theories were counter-examples to its generality. As a consequence,
similar doubts are casted on the results of [1]. The question which
arises in this context is how is it possible that the result
described in [1] is at variance with the simple fact that a causal
theory already exists? In the following, the proof of the non
existence of deterministic hidden variable theories is critically
commented on by stressing some failures in the reasoning leading to
this conclusion. More precisely we will see that the author of [1] ,
by limiting his discussion to a very limited class of hidden
variable theories, namely the class of the non contextual
theories[5], invalidates necessarily the generality of his
conclusions.

In order to analyze the possible existence of local hidden
variables, as done by Bell [2], we introduce a probability density
$\rho\left(\lambda\right)$ running on the dynamical space of the
hidden variables $\lambda$. These variables label all trajectories
and they could be for example the initial parameters determining the
motion of the photon. Here, without loss of generality, we limit our
analysis to local and deterministic hidden variables models in which
we suppose non stochastic elements at the fundamental level. The
generalization to objective local theories [11] is straightforward
and is not necessary for the present purpose. In the case of two
separated particles with spins, Bell introduced the definition
\begin{equation}
E_{c}\left(\mathbf{a},\mathbf{b}\right)= \int d\lambda
\rho\left(\lambda\right)
A\left(\mathbf{a},\lambda\right)B\left(\mathbf{b},\lambda\right)\label{5}
\end{equation}
 where $A=\pm 1$ and $B=\pm 1$ are observable associated with the
 polarization of the two photons measured along the axis $\mathbf{a}$ and $\mathbf{b}$ labeled by the rotator angles $\theta_{A}$ and $\theta_{B}$.
 This ``classical''
expectation value $E_{c}$ is anticipated to be equivalent to the one
predicted by quantum mechanics for correlated events and it is with
such a definition that Bell proved a contradiction between the two
following hypothesis: 1) strict validity of quantum mechanics, and:
2) existence of a subjacent and hidden local dynamical structure.\\
The locality assumption
 used here supposes that A does not depend on  $\mathbf{a}$
 and that similarly  B does not depend on  $\mathbf{b}$. This is just a
  mathematical formulation of a famous sentence written in the
original Einstein-Podolsky-Rosen (EPR) article [12]: ``since at the
time of measurement the two
  systems no longer interact, no real change can take place in the
second system in consequence of anything that may  be done to the
first system''.
In this model it is easy to prove that the Bell
number
\begin{eqnarray}
S_{c}=E_{c}\left(\theta_{A},\theta_{B}\right)+E_{c}\left(\theta_{A'},\theta_{B'}\right)
\nonumber\\
+E_{c}\left(\theta_{A'},\theta_{B}\right)-E_{c}\left(\theta_{A},\theta_{B'}\right)
\end{eqnarray}
which can be written
\begin{eqnarray}
S_{c}=\int[A\left(\mathbf{a},\lambda\right)
\left(B\left(\mathbf{b},\lambda\right)-B\left(\mathbf{b'},\lambda\right)\right)\nonumber\\+A\left(\mathbf{a'},\lambda\right)
\left(B\left(\mathbf{b},\lambda\right)+B\left(\mathbf{b'},\lambda\right)\right)]\rho\left(\lambda\right)d\lambda
\end{eqnarray}
 must obey the Bell inequality
\begin{equation}
-2 \leq S_{c}\leq 2.\end{equation} In the Bohm-EPR [13] experiment
studied by Bell, the locality assumption is transformed
mathematically into a criterion for the independency of measurements
made on two entangled photons separated by an arbitrary large
distance. This formula has been applied in Ref.[1] by considering
that the two measurements made on the same photon are independent in
the experiment. But this axiom is not acceptable because the two
splits of the photon wave packets are separated in time and this
must introduce some strong correlation between these successive
events. It is then clear that the order of operation is crucial in
the experiment. Therefore, an expression such as Eq.~1 cannot be
written because the result obtained in
`B' or `C' must depend on the result obtained in `A'.\\
 More precisely, if one accepts that the probabilities offered by the quantum
mechanical prediction can be explained in terms of an average over hidden variables, then the relation
$P\left(\beta,\alpha,\theta_{B},\theta_{A}\right)=\int
dP\left(\beta,\alpha,\lambda,\theta_{B},\theta_{A}\right)$ holds, where the
observables $A$ and $B$ take the values $\alpha$ and $\beta$, respectively. Using
a conditional probability,
$dP\left(\beta,\alpha,\lambda,\theta_{B},\theta_{A}\right)$ can be transformed into
\begin{eqnarray}
P_{2}\left(\beta|\alpha,\lambda,\theta_{B},\theta_{A}\right)
dP\left(\alpha,\lambda,\theta_{B},\theta_{A}\right)
\nonumber\\=P_{2}\left(\beta|\alpha,\lambda,\theta_{B},\theta_{A}\right)P_{1}\left(\alpha|\lambda,\theta_{A}\right)
\rho\left(\lambda\right)d\lambda.\label{proba1}
\end{eqnarray} Here the
conditional probabilities
$P_{2}\left(\beta|\alpha,\lambda,\theta_{B},\theta_{A}\right)$ and
$P_{1}\left(\alpha|\lambda,\theta_{A}\right)$, which depend on the
complete history of the particle, can only take the values 0 or 1.
This is due to the fact that in the context of deterministic hidden
variable models a particle with a given $\lambda$ can only follow a
single path. For the same reason, we have
$P_{1}\left(+1|\lambda,\theta_{A}\right)+P_{1}\left(-1|\lambda,\theta_{A}\right)=1$
and, identically,
$\sum_{\alpha,\beta}P_{2}\left(\beta|\alpha,\lambda,\theta_{B},\theta_{A}\right)=1$.
In this notation, the relations
$P_{1}\left(\alpha|\lambda,\theta_{B},\theta_{A}\right)=P_{1}\left(\alpha|\lambda,\theta_{A}\right)$
and
$\rho\left(\lambda,\theta_{B},\theta_{A}\right)=\rho\left(\lambda\right)
$ have been implicitly supposed to be true. This is based on a
locality assumption. The mean value of $AB$ is defined by
\begin{eqnarray}
E\left(\theta_{A},\theta_{B}\right)=\sum_{\alpha,\beta}\alpha\beta
\cdot
P\left(\beta,\alpha,\lambda,\theta_{B},\theta_{A}\right)\nonumber\\
=\int\sum_{\alpha,\beta}\alpha\beta\cdot
P_{2}\left(\beta|\alpha,\lambda,\theta_{B},\theta_{A}\right)P_{1}\left(\alpha|\lambda,\theta_{A}\right)
\rho\left(\lambda\right)d\lambda.\label{proba2}
\end{eqnarray}
 In the case of two independent measurements, as
considered by Bell, it is permitted to write
\begin{eqnarray}
E\left(\theta_{A},\theta_{B}\right)\nonumber\\=\int\sum_{\beta}\beta\cdot
P_{2}\left(\beta|\lambda,\theta_{B}\right)\sum_{\alpha}\alpha\cdot
P_{1}\left(\alpha|\lambda,\theta_{A}\right)
\rho\left(\lambda\right)d\lambda.
\end{eqnarray}
This is just an other formulation of Eq.~4 and
\begin{eqnarray}
A\left(\mathbf{a},\lambda\right)=\sum_{\alpha}\alpha\cdot
P_{1}\left(\alpha|\lambda,\theta_{A}\right)\nonumber\\B\left(\mathbf{b},\lambda\right)=\sum_{\beta}\beta\cdot
P_{2}\left(\beta|\lambda,\theta_{B}\right)\label{proba3}
\end{eqnarray}
are only allowed to take the values $\pm1$ as expected, because the
sum contains only one non-zero term. However, this factorization,
corresponding to the case of the EPR situation, cannot apriori be
accepted in the Gedanken experiment of Ref. [1] and the complete
formula~5 must be considered. It is still possible to write
\begin{eqnarray}
A\left(\mathbf{a},\lambda\right)=\sum_{\alpha}\alpha\cdot
P_{1}\left(\alpha|\lambda,\theta_{A}\right)
\end{eqnarray} but now
\begin{eqnarray}
B\left(\mathbf{b},\mathbf{a},\lambda\right)=\sum_{\alpha,\beta}\beta\cdot
P_{2}\left(\beta|\alpha,\lambda,\theta_{B},\theta_{A}\right)P_{1}\left(\alpha|\lambda,\theta_{A}\right)
\end{eqnarray} and
\begin{eqnarray}
AB\left(\mathbf{b},\mathbf{a},\lambda\right)=\sum_{\alpha,\beta}\alpha\beta\cdot
P_{2}\left(\beta|\alpha,\lambda,\theta_{B},\theta_{A}\right)
P_{1}\left(\alpha|\lambda,\theta_{A}\right).\nonumber\\
\end{eqnarray}
In addition, one has
$AB\left(\mathbf{b},\mathbf{a},\lambda\right)=A\left(\mathbf{a},\lambda\right)\cdot
B\left(\mathbf{b},\mathbf{a},\lambda\right)$ as seen once again from
the fact that for a given $\lambda$ the double sum contains only one
non-zero term. This analysis proves clearly that formula 4 is not
adapted to the situation described in Ref.[1]. As a consequence, the
deduction of Ref.[1] is erroneous in the absence of additional
elements allowing for a choice between Eq.~8 and Eqs.~9-11.\\

However, it is easy to build a model in which Eq.~8 (i.~e.~Eq.~1) is
wrong and Eqs.~9-11 are simultaneously true. As explained at the
beginning we consider indeed the de Broglie-Bohm [8,10] mechanics as
a counter-example of the hidden variable model which can describe
such an experiment in terms of deterministic trajectories.
Originally the model of de Broglie-Bohm is defined in the case of
the non relativistic Schrodinger equation because it is only in this
case that a simple generalization for the many-body problem is
obtained [10]. In such theory particles are point like objects
moving together with a guiding wave solution of the ordinary
Schrodinger equation in the configuration space. Because the
reasoning of Ref.[1] is general, we could in fact think in terms of
neutrons passing through Stern and Gerlach beam splitters and we
would have directly a counter example to the argumentation of [1].
However, this is not necessary because generalizations of the de
Broglie-Bohm initial model have been proposed by Bohm himself [10]
and others [14-17] based on the quantum wave functional formalism.
In such model the photon field is a each instant defined by a
function $A_{\mu}\left(x\right)$ obeying to some generalization of
the Maxwell equations taking into account a generally highly
nonlocal quantum potential $Q[A_{\mu}\left(x\right)]$. We can in
analogy with the initial model proposed by de Broglie define a
guidance formula for the field and justify, in principle, the
probabilistic predictions of quantum optics. In the experiment
discussed in [1] the field is divided at each polarizing beam
splitters (PBS) and the interactions with fermionic and localized
detectors explain well known rules of photo counting. This is
directly in conflict with [1]
prohibiting any causal explanations of that kind.\\
 It can be
observed that at the single particle level a more intuitive model is
possible [18-20]: by analogy with the de Broglie dynamics [8,10],
the photon is supposed to be a particle guided by an electromagnetic
wave and its velocity $d\boldsymbol{\lambda}\left(t\right)/dt$ can
be defined by the velocity of energy in the guiding wave. In the
present situation the incident guiding plane wave before the PBS is
characterized by its linear polarization $\theta$. After crossing
the PBS the plane wave is split into two components and the
probability for a photon to be in one beam or in the other is given
by Malus law. However, depending on its initial position before the
PBS, the photon will follow a particular trajectory to go to only
one of the two possible exits. This is sufficient to explain the
result of the experiment proposed in [1] in a completely
deterministic way since the energy flow
 is obviously well defined with the Maxwell equations and must depends on the chronology of the operations at the two beam splitters.\\
Going back now to the general discussion we see that formula 1 and 8
cannot be used in the hidden variable models presented here without
violating the Bell inequality which contradicts the fact that the
model are completely causal.  This implies that $B$ must depend on
the two orientations $\mathbf{a}$ and $\mathbf{b}$. Due to the flow
continuity, this is equivalent to say that the hidden variable
models should be highly contextual in order to justify the
experimental results. This fact is well known since the works of
Kochen, Specker and Bell [5,22] and from the analysis of de Broglie
and Bohm [8,10].
\section{Conclusion} To conclude, the author of [1]
considers only a very restrictive class of hidden variable theories
in which the measurements have no back action on the hidden
variable. He can not consequently recuse all hidden parameters
models. We observe that a ``gedanken'' experiment as the one
proposed in [1] can be explained in terms of deterministic theories
``\`{a} la de Broglie'' [20] or ``\`{a} la Bohm'' [17]. Such models
are highly contextual [22] and consequently the dynamic of the
hidden parameters depends necessarily on the order of the
measurements. However, one must be aware of the fact that such
deterministic models do modify the classical intuition coming from
Newtonian dynamics (see for example [24,25]). Serious problems do
occur in particular when the concept of entanglement is considered
in the many-body situation. Only in these cases must the classical
determinism should be necessarily revised with the abandon of the
locality conditions. This fact, which was implicit in the work of
Einstein Podolsky and Rosen, and which was rigourously proven later
by Bohm and Bell, does not sign up the end of the classical
determinism of Laplace.
\section{Acknowledgements}
I would like to thank Serge Huant for helpful comments and
discussions, and I would like to thanks A.~Zeilinger and his group
for supports during the year 2003 in which the first ideas of this
work were conceived.

\end{document}